\documentstyle[11pt,epsf]{article}
\font\elevenbf=cmbx10 scaled\magstep 1
\font\elevenrm=cmr10 scaled\magstep 1
 1

\renewenvironment{thebibliography}[1]
 { \elevenrm
   \begin{list}{\arabic{enumi}.}
    {\usecounter{enumi} \setlength{\parsep}{0pt}
     \setlength{\itemsep}{3pt} \settowidth{\labelwidth}{#1.}
     \sloppy
    }}{\end{list}}
\renewcommand{\thefootnote}{\fnsymbol{footnote}}
\begin{document}
\noindent
\thispagestyle{empty}
\renewcommand{\thefootnote}{\fnsymbol{footnote}}
\begin{flushright}
{\bf TTP95-10}\footnote{The complete postscript file of this
preprint, including figures, is available via anonymous ftp at
ttpux2.physik.uni-karlsruhe.de (129.13.102.139) as
/ttp95-10/ttp95-10.ps or via www at
http://ttpux2.physik.uni-karlsruhe.de/cgi-bin/preprints/
Report-no: TTP95-10.}\\
{\bf March 1995}\\
hep-ph/9505262 \\
\end{flushright}
\begin{center}
  \begin{large}
 RADIATION OF LIGHT FERMIONS IN HEAVY FERMION PRODUCTION\footnote{work
supported by BMFT 056 KA 93P}\\
  \end{large}
  \vspace{0.5cm}
  \begin{large}
   A.H. Hoang, J.H. K\"uhn and
   T. Teubner\footnote{e--mail: tt@ttpux2.physik.uni-karlsruhe.de}
 \\
  \end{large}
  \vspace{0.1cm}
Institut f\"ur Theoretische Teilchenphysik, Universit\"at Karlsruhe, \\
D--76128 Karlsruhe, Germany \\
  \vspace{0.7cm}
  {\bf Abstract}\\
\vspace{0.3cm}
\renewcommand{\thefootnote}{\arabic{footnote}}
\addtocounter{footnote}{-2}
\noindent
\begin{minipage}{15.0cm}
\begin{small}
The rate for the production of a pair of massive fermions in $e^+ e^-$
annihilation plus real or virtual radiation of a pair of massless
fermions is calculated analytically.
The contributions for real and virtual radiation are displayed separately.
The asymptotic behaviour close to threshold and for high energies
is given in a compact form. These approximations provide
arguments for the appropriate choice of the scale in
the ${\cal O}(\alpha)$ result, such that no large logarithms remain in the
final answer.
\end{small}
\end{minipage}
\end{center}
\vspace{1.2cm}
%
%
%
\noindent
{\em 1.~Introduction}\\
QED reactions leading to four fermion final states
$f_1 \bar f_1 f_2 \bar f_2$ with fermion masses $m_1$ and $m_2$ have
been considered in the literature a long time ago~\cite{BFK} and the
final result was at that time expressed in the form of a two
dimensional integral. The cross sections were calculated for leptons
in the context of QED. However, it is fairly evident
that a large part of the results can easily be transferred to the
corresponding
QCD reactions. The analogy becomes even closer when considering the mass
assignments of interest for actual physical reactions: either
equal masses $m_1 = m_2$ or alternatively $m_1 \gg m_2$.
This mass hierarchy applies
equally well to leptons and to quarks and hence to QCD calculations and
will be exploited in this work.
\par
Apart of the ``exclusive'' channel, where all four fermions are detected,
also the inclusive rate for $f_1 \bar f_1 + anything$ is of practical
interest. To ${\cal O}(\alpha)$ this calculation has been presented in the
classic book by Schwinger~\cite{Schwinger}. Higher order results are
available in the context of QCD in the limit of vanishing fermion mass
\cite{as2,as3} or including mass corrections
through an expansion in $m^2/s$. Terms of order $m^2/s$ where calculated
up to $\alpha_s^3$ \cite{CKfirst}, the $m^4/s^2$ terms up
to $\alpha_s^2$ \cite{CKlast}.
This expansion is, however, inadequate close to threshold. In this
kinematical region the full calculation up to order $\alpha_s^2$ would
be required. This is particularly desirable in view of the fact that
the ambiguity in the scale $\mu^2$ in $\alpha_s$ leads to a large
uncertainty in the leading order correction. Close to threshold both
the mass and the three-momentum of the produced fermions seem to be
reasonable choices for $\mu^2$, giving rise, however, to drastically
different predictions.
\par
The analytical calculation of the cross section, including the full mass
and energy dependence and all (real and virtual) gauge boson contributions
seems like a difficult task. However, the subclass
of diagrams involving the real and virtual radiation of light fermions
is more accessible. In this paper the result for both real and virtual
radiation will be presented for arbitrary $m_1^2$ and $s$, in the limit
$m_1^2 \gg m_2^2$.
Combining the two contributions, one arrives at a result which still
exhibits mass singularities of the form $\ln(m_2^2/s)$. They can be
removed by adopting the $\overline{\mbox{MS}}$ scheme.
Normalizing the coupling constant at scale $\mu^2 = s$ eliminates
all large logarithms, at least away from the threshold region.
This provides the first step in the calculation of
corrections to order $\alpha_s^2$.
The relatively compact analytical result
can then be studied in the limit close to threshold as well as in the
high energy region.
\par
\vspace{1cm}\noindent
{\em 2.~Real radiation}\\
For definiteness the cross section for the production of the
$f_1 \bar f_1 f_2 \bar f_2$ final state
in $e^+ e^-$ annihilation through a virtual photon will
be considered, normalized relatively to the point cross section.
The result is evidently also applicable to Z decays through the
vector current.
\begin{figure}[htb]
\begin{center}
\leavevmode
\epsfxsize=5cm
\epsffile[230 380 370 450]{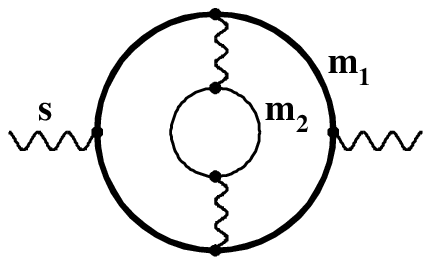}
\leavevmode
\epsfxsize=5cm
\epsffile[230 380 370 450]{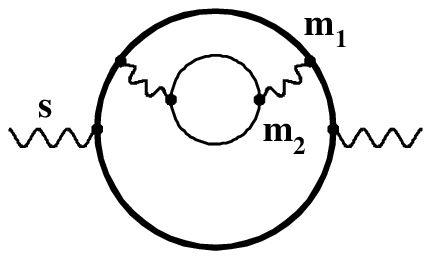}\\
\vskip 5mm
\caption{\label{fig1} {\em Characteristic Feynman diagrams describing
the production of a pair of massive and a (real or virtual) pair of
massless fermions.}}
\end{center}
\end{figure}
\noindent
The relevant Feynman amplitudes can be derived from the four fermion cuts
of the diagrams with two closed fermion loops, with representative
examples depicted in Fig.~\ref{fig1}.
The rate can be expressed by a two dimensional integral~\cite{BFK}:
\begin{eqnarray}
\label{masterreal}
\lefteqn{
R_{f_1\bar f_1 f_2\bar f_2} \equiv
\frac{\sigma_{f_1 \bar f_1 f_2 \bar f_2}}
     {\sigma_{f_1 \bar f_1,pt}} =
   \left(\frac{\alpha}{\pi}\right)^2\,\varrho^R \,, }
  \\[2mm]
\lefteqn{
\varrho^R = \frac{4}{3}\,
 \int\limits_{4m_1^2/s}^{(1-2m_2/\sqrt{s})^2} {\rm d}y
 \int\limits_{4m_2^2/s}^{(1-\sqrt{y})^2} \frac{{\rm d}z}{z} \,
  \left(1 + \frac{2m_2^2}{s\,z}\right)\,
  \sqrt{1 - \frac{4m_2^2}{s\,z}} \, \times
 } \nonumber \\[1mm]
 & & \left\{
 \frac{\frac{2m_1^4}{s^2}+\frac{m_1^2}{s}(1-y+z)-\frac{1}{4}
           (1-y+z)^2-\frac{1}{2}(1+z)y}{1-y+z}\,
  \ln\frac{1-y+z-\sqrt{1-\frac{4m_1^2}{s\,y}}\,\Lambda^{1/2}(1,y,z)}
          {1-y+z+\sqrt{1-\frac{4m_1^2}{s\,y}}\,\Lambda^{1/2}(1,y,z)}
 \,\right.\nonumber\\[1mm]  & & \left.
 \mbox{} \, - \,
 \sqrt{1-\frac{4m_1^2}{s\,y}} \, \Lambda^{1/2}(1,y,z)\,\left[ \,
  \frac{1}{4} + \frac{\frac{2m_1^2}{s}+\frac{4m_1^4}{s^2}+
                      \left(1+\frac{2m_1^2}{s}\right)z}
                     {(1-y+z)^2 -
                      \left(1-\frac{4m_1^2}{s\,y}\right) \, \Lambda(1,y,z)}
  \, \right]\,  \right\}
\end{eqnarray}
where
\begin{equation}
\Lambda(1,y,z)=1+y^2+z^2-2(y+z+y\,z) \,.\nonumber
\end{equation}
For arbitrary $m_1$, $m_2$ and $s$ already the first integration
leads to elliptic functions and fairly lengthy expressions.
However, in the limit $m_1^2 \gg m_2^2$ the radiation can be split
into two parts: ``soft'' radiation with energy of the $f_2 \bar f_2$
system smaller than a cutoff $\Delta$, with $m_2 \ll \Delta \ll m_1$,
and the remainder, denoted ``hard'' radiation. The two parts can be
integrated separately, and their sum is given by
\begin{equation}
\label{fullreal}
\rho^R =
           f_{R}^{(2)}\,\ln^2\frac{m_2^2}{s} +
           f_{R}^{(1)}\,\ln\frac{m_2^2}{s} +
           f_{R}^{(0)} \,.
\end{equation}
As expected, the cutoff $\Delta$ cancels in the sum.
The functions multiplying the second and first power of the logarithm
are closely related to
the well-known Schwinger result \cite{Schwinger}
for real radiation of a light vector particle with mass
$\lambda\ll m_1$:
\begin{equation}
\label{onetworeal}
\frac{\sigma_{f_1 \bar f_1 \gamma}}{\sigma_{f_1 \bar f_1, pt}} =
 6\,\bigg(\frac{\alpha}{\pi}\bigg) \, \bigg[\, f_R^{(2)} \, \bigg(
 \ln \frac{s}{\lambda^2} + \frac{5}{3} \, \bigg) - \frac{1}{2} \, f_R^{(1)}
 \,\bigg]\,.
\end{equation}
The evaluation of the function without logarithmic
enhancement constitutes the main effort of this section. The three
functions $f_{R}^{(0,1,2)}$ are given by
\begin{eqnarray}
f_{R}^{(2)} & = & - \frac{3 - w^2}{24}\,
  \left[ \, \big( 1 + {w^2} \big) \,\ln p + 2\,w \, \right]
\,, \\[2mm]
f_{R}^{(1)} & = &
- \frac{\left( 3 - {w^2} \right) \,\left( 1 + {w^2} \right) }{6}\,
     \bigg[ \mbox{Li}_2(p) + \mbox{Li}_2({p^2})\,\nonumber\,
        \\
 & & \quad \mbox{} + \frac{1}{4}\,
        \bigg( 4\,\ln w + 5\,\ln p - 6\,\ln({{1 - {w^2}}\over 4}) \bigg) \,
        \ln p - 2\,\zeta(2) \, \bigg] \,\nonumber\,\\
 & & \mbox{} -
  \frac{w\,\left( 3 - {w^2} \right) }{3}\,
   \bigg[ \, \ln({{1 - {w^2}}\over 4}) - 2\,\ln w \, \bigg]  +
  \frac{39 - 70\,{w^2} + 23\,{w^4}}{144}\,\ln p \nonumber\,\\
 & & \mbox{} +
  \frac{w\,\left( -177 + 71\,{w^2} \right) }{72}
\,,\\[2mm]
f_{R}^{(0)} & = &
\frac{\left( 3 - {w^2} \right) \,\left( 1 + {w^2} \right) }{6}\,
   \bigg[ \,4\,\mbox{Li}_3(1 - p) + 3\,\mbox{Li}_3({p^2}) +
     4\,\mbox{Li}_3({p\over {1 + p}}) + 5\,\mbox{Li}_3(1 - {p^2}) -
     \frac{13}{2}\,\zeta(3)\,\nonumber\,\\
 & & \mbox{}\,\quad +
     2\,\ln({{4\,{w^2}}\over {1 - {w^2}}})\,
      \Big( \mbox{Li}_2(p) + \mbox{Li}_2({p^2}) \Big)  +
     \bigg( 3\,\ln({{1 - {w^2}}\over 4}) - 4\,\ln p - 4\,\ln({w^2}) \bigg)\,
      \zeta(2)\,\nonumber\,\\
 & & \mbox{}\,\quad +
     \frac{1}{12}\,\ln^3({{1 - {w^2}}\over 4}) +
     \frac{1}{12}\,\ln p \,\bigg( 51\,\ln^2({{1 - {w^2}}\over 4}) +
        46\,\ln^2 p - 99\,\ln p \,\ln({{1 - {w^2}}\over 4}) \bigg) \,
      \nonumber\,\\
 & & \mbox{}\,\quad +
     \ln w \,\ln p \,\bigg( 8\,\ln w + 11\,\ln p -
        12\,\ln({{1 - {w^2}}\over 4}) \bigg) \,\bigg] \,\nonumber\,
   \\
 & & \mbox{} + \frac{
    \left( 1 - {w^2} \right) \,\left( 21 - 13\,{w^2} \right) }{24}\,
   \mbox{Li}_2(p) + \frac{-15 - 58\,{w^2} + 17\,{w^4}}{72}\,
   \Big( \mbox{Li}_2(p) + \mbox{Li}_2({p^2}) \Big) \,\nonumber\,
   \\
 & & \mbox{} + \frac{2\,w\,\left( 3 - {w^2} \right) }{6}\,
   \bigg( -\ln^2({{4\,{w^2}}\over {1 - {w^2}}}) + 3\,\zeta(2) \bigg)  +
  \frac{-33 + 218\,{w^2} - 73\,{w^4}}{72}\,\zeta(2)\,\nonumber\,
   \\
 & & \mbox{} + \frac{-27 - 3\,w + 72\,{w^2} - 98\,{w^3} - 33\,{w^4} +
     31\,{w^5}}{72\,w}\,\ln^2 p \,\nonumber\,\\
 & & \mbox{} +
  \bigg( \frac{-15 - 34\,{w^2} + 11\,{w^4}}{18}\,\ln w +
     \frac{15 + 42\,{w^2} - 13\,{w^4}}{24}\,\ln({{1 - {w^2}}\over 4}) \bigg)
    \,\ln p \,\nonumber\,\\
 & & \mbox{} +
  \frac{2451 - 766\,{w^2} - 205\,{w^4}}{864}\,\ln p +
  \frac{w\,\left( 177 - 71\,{w^2} \right) }{36}\,
   \ln({{4\,{w^2}}\over {1 - {w^2}}})\,\nonumber\,\\
 & & \mbox{} +
  \frac{w\,\left( -2733 + 1067\,{w^2} \right) }{432}
      \,
\end{eqnarray}
where
\begin{equation}
\qquad w \equiv \sqrt{1-4m_1^2/s} \,\,, \qquad
       p \equiv \frac{1-w}{1+w}\,.
\end{equation}
$\mbox{Li}_2$, $\mbox{Li}_3$ denote the di- and trilogarithms,
$\zeta(2)$ and $\zeta(3)$ the Zeta function of the respective
arguments~\cite{Lewin}.
The details of this calculation will be given
elsewhere~\cite{HKTfuture}.
This result is directly applicable for example to the annihilation
of $e^+ e^-$ into $\tau^+ \tau^-$ with the additional ``final state''
radiation of $e^+ e^-$ (or $\mu^+ \mu^-$) which adds incoherently
to the process with $\tau$ radiated from the light fermion final
state~\cite{HJKT2}.
The behaviour close to threshold ($w \to 0$) and in the
high energy region ($m_1^2/s \to 0$) is of relevance for direct
applications and for the discussion of our results presented below.
One finds
\begin{eqnarray}
\varrho^R &  \stackrel{w\to 0}{\longrightarrow} &
{w^3}\,\bigg[\, \frac{1}{3}\,\ln^2 \frac{m_2^2}{s} -
    \frac{4}{9}\,\ln \frac{m_2^2}{s}\,\Big( 6\,\ln(2\,w) - 13 \Big) \,
     \nonumber\,\\
 & & \mbox{}\,\qquad\,\qquad +
    \frac{16}{3}\,\ln^2(2\,w) - \frac{208}{9}\,\ln(2\,w) - 4\,\zeta(2) +
    \frac{898}{27} \,\bigg] \,+{\cal O}(w^5)\,,
    \label{realthresh} \\[2mm]
\varrho^R &  \stackrel{\frac{m_1^2}{s}\to 0}{\longrightarrow} &
-\, \frac{1}{6}\,\ln^2 \frac{m_2^2}{s}\,\Big( \ln x + 1 \Big) +
  \ln \frac{m_2^2}{s}\,\bigg( \frac{1}{6}\,\ln^2 x - \frac{13}{18}\,\ln x +
     \frac{4}{3}\,\zeta(2) - \frac{53}{36} \bigg) \,\nonumber\,
   \\
 & & \mbox{}\,\quad - \frac{1}{18}\,\ln^3 x + \frac{13}{36}\,\ln^2 x -
  \bigg( \frac{133}{108} + \frac{2}{3}\,\zeta(2) \bigg) \,\ln x +
  \frac{5}{3}\,\zeta(3) + \frac{32}{9}\,\zeta(2) -
  \frac{833}{216}\,\nonumber\,\\
 & & \mbox{} +
  x\,\bigg[\, -\,\frac{1}{3}\,\ln^2 \frac{m_2^2}{s}   +
     \ln \frac{m_2^2}{s}\,\left( \frac{2}{3}\,\ln x - \frac{40}{9} \right) \,
      \nonumber\,\\
 & & \mbox{}\,\qquad\,\qquad -
     \frac{1}{3}\,\ln^2 x + \frac{13}{9}\,\ln x - \frac{4}{3}\,\zeta(2) -
     \frac{233}{27} \,\bigg] \,
     +{\cal O}(x^2)
     \label{realhigh}
\end{eqnarray}
where
\begin{equation}
x\equiv\frac{m_1^2}{s}\,.
\end{equation}
Close to threshold radiation of light fermions is strongly suppressed,
similar to the radiation of photons as calculated in ${\cal O}(\alpha)$.
The leading logarithms in the high energy limit coincide with those
given in~\cite{BFK}. The prediction for
$R_{f_1 \bar f_1 f_2 \bar f_2}$ (with $m_1/m_2 = m_\tau/m_e = 3477.5$
chosen for
illustrative purpose) based on eq.~(\ref{masterreal}),
is shown in Fig.~\ref{real} (solid line).
Also shown are the high energy approximation, eq.~(\ref{realhigh}),
(dashed line) including the
linear term in $x$ and the threshold approximation, eq.~(\ref{realthresh}),
(dashed-dotted line). It is evident that the
high energy approximation can be used for values of $x=m_1^2/s$ between
$0$ and about $0.125$ corresponding to values of $w$ from $1$ down
to $0.7$ and hence surprisingly close to the threshold. For $w$ below
this value the threshold approximation provides an adequate description.
\begin{figure}[htb]
\begin{center}
\leavevmode
\epsfxsize=5cm
\epsffile[220 420 420 550]{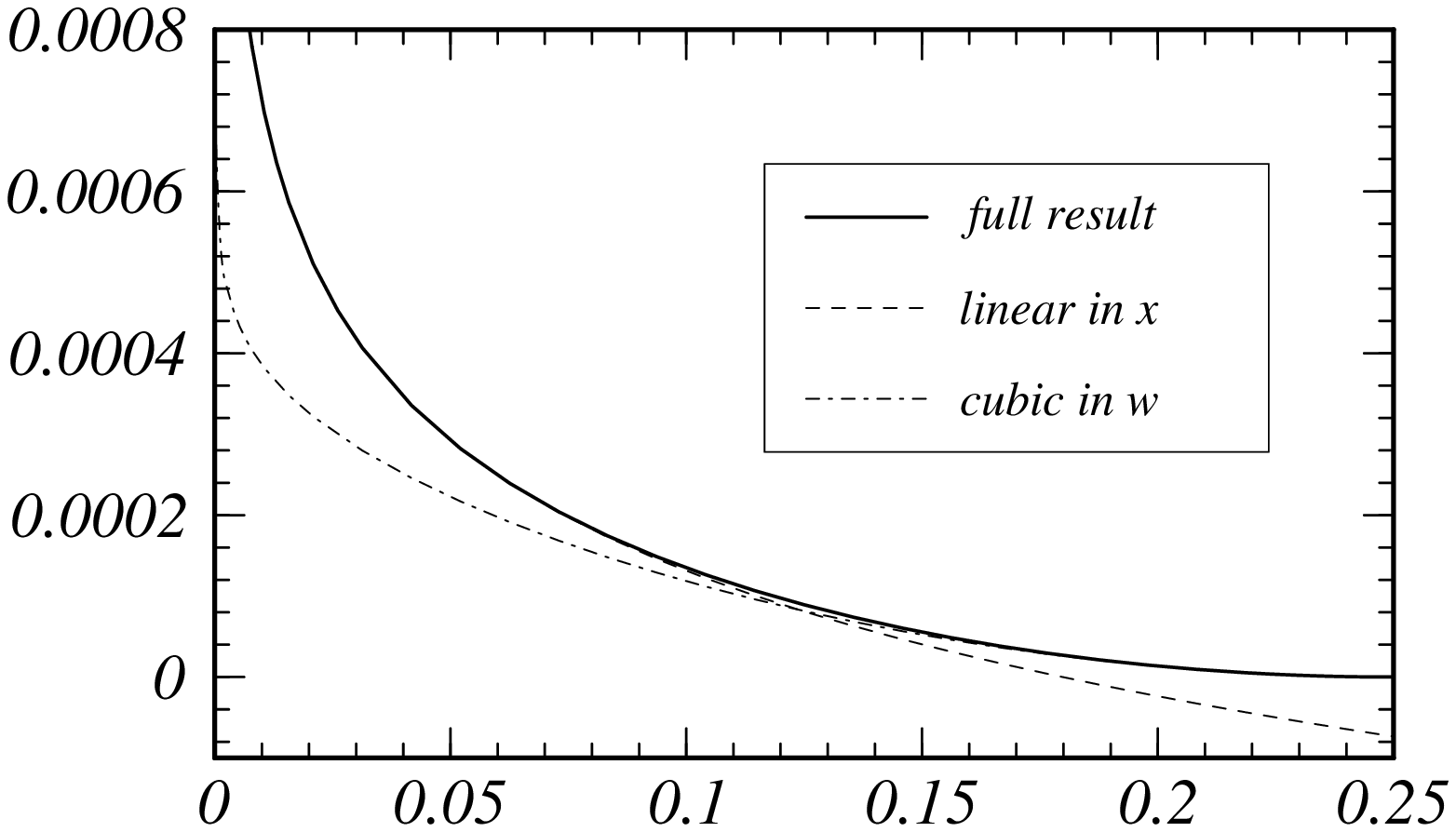}
\vskip 30mm  $\displaystyle{\mbox{\hspace{12.5cm}}\bf\frac{m_1^2}{s}}$
\vskip 5mm
\caption{\label{real} {\em
Production rate $R_{f_1 \bar f_1 f_2 \bar f_2}$ based on the exact result
(solid line) and approximations described in the text as functions of
$m_1^2/s$ with $m_1 = m_{\tau}$ and $m_2 = m_e$.}}
\end{center}
\end{figure}
\pagebreak[4]
\par
\vspace{1cm}
\noindent
{\em 3.~Virtual corrections}\\
Virtual corrections in the present context arise from the two
particle cut of the ``double bubble'' diagrams (Fig.~\ref{fig1}).
The ${\cal O}(\alpha^2)$ corrections $\delta\Lambda_\mu$ to the
lowest order vertex can be classified into contributions to the
Dirac ($F_1$) and the Pauli form factor ($F_2$)
\begin{eqnarray}
\label{deltalambda}
\delta\Lambda_\mu =
\gamma_\mu\,\left(\frac{\alpha}{\pi}\right)^2\,F_1 +
  \frac{i}{2\,m_1}\,\sigma_{\mu\nu}\,q^\nu\,
  \left(\frac{\alpha}{\pi}\right)^2\,F_2
\end{eqnarray}
where
$\sigma_{\mu\nu} = \frac{i}{2}[\gamma_\mu,\gamma_\nu]$
and $q$ denotes the photon momentum flowing into the vertex.
Using the dispersive methods applied already in~\cite{KKKS,HJKT1}
the calculation of $F_{1,2}$ can be easily reduced to a one dimensional
integration. It results from the convolution of the massive vector boson
exchange vertex correction to $f_1 \bar f_1$ production with the
absorptive part of the vacuum polarization of fermions with mass
$m_2$. Denoting the vector boson mass by $\lambda$ the
convolution reads
\begin{eqnarray}
\label{mastervirt}
F_{1,2} & = & \frac{1}{3}\,\int_{4\,m_2^2}^\infty
   \frac{{\rm d}\lambda^2}{\lambda^2}\,
   \left(1+\frac{2\,m_2^2}{\lambda^2}\right)\,
   \sqrt{1-\frac{4\,m_2^2}{\lambda^2}}\,
   \widehat{F}_{1,2}(\lambda^2)
\end{eqnarray}
with
\begin{eqnarray}
\mbox{Re}\,\widehat{F}_1(\lambda^2) & = &
- \frac{1}{4\,w}\,\bigg[ 1 + {w^2} +
       \frac{1 - {w^2}}{{w^2}}\,\frac{\lambda^2}{m_1^2}\,
        \bigg( 1 + \frac{
            \left( 3 - 2\,{w^2} \right) \,
             \left( 1 - {w^2} \right) }{8\,{w^2}}\,
           \frac{\lambda^2}{m_1^2} \bigg) \, \bigg] \,
     \Psi(p,\frac{\lambda^2}{m_1^2})\,\nonumber\,
     \\
 & & \mbox{}  +
  \bigg[ \frac{-1 + 2\,{w^2}}{2\,{w^2}} +
     \frac{-3 + 7\,{w^2} + 2\,{w^4}}{8\,{w^4}}\,
      \frac{\lambda^2}{m_1^2} +
     {3\over {\frac{\lambda^2}{m_1^2} - 4}} \,\bigg] \,
   \Phi(\frac{\lambda^2}{m_1^2})\,\nonumber\,\\
 & & \mbox{} +
   \frac{1}{16\,{w^4}}\,\bigg[ -3 + 7\,{w^2} + 2\,{w^4} +
     \frac{{{\left( 1 - {w^2} \right) }^2}\,
        \left( -3 + 2\,{w^2} \right) }{2\,w}\,\ln p \,\bigg]
    \,\frac{\lambda^4}{m_1^4}\,\ln \frac{\lambda^2}{m_1^2}\,
   \nonumber\,\\
 & & \mbox{} -
  \frac{1}{2\,{w^2}}\,\bigg[ 1 +
     \frac{1 - {w^2}}{2\,w}\,\ln p \,\bigg] \,
   \frac{\lambda^2}{m_1^2}\,\ln \frac{\lambda^2}{m_1^2} -
  \bigg[ \frac{1}{2} + \frac{1 + {w^2}}{4\,w}\,\ln p \,\bigg]
    \,\ln \frac{\lambda^2}{m_1^2}\,\nonumber\,\\
 & & \mbox{}
   + \bigg[\frac{{{\left( 1 - {w^2} \right) }^2}\,
     \left( -3 + 2\,{w^2} \right) }{64\,{w^5}}\,\ln^2 p\,\bigg]\,
   \frac{\lambda^4}{m_1^4}\,\nonumber\,\\
 & & \mbox{} +
  \frac{1}{4\,{w^2}}\,\bigg[ -\left( 1 + 2\,{w^2} \right)  +
     \frac{1 - {w^2}}{2\,w}\,\ln p\,
      \Big( -3 + 2\,{w^2} - \ln p \Big) \, \bigg] \,
   \frac{\lambda^2}{m_1^2}\,\nonumber\,\\
 & & \mbox{} -
  \frac{1}{8\,w}\,\ln p\,
   \bigg( 2\,( 1 + 2\,{w^2} )  +
     ( 1 + {w^2} ) \,\ln p \bigg)  - 1 \, ,
   \label{F1oneloop}\\[2mm]
\mbox{Re}\,\widehat{F}_2(\lambda^2) & = &
\frac{{{\left( 1 - {w^2} \right) }^2}}{4\,{w^3}}\,
   \bigg[ 1 + \frac{3\,\left( 1 - {w^2} \right) }{8\,{w^2}}\,
      \frac{\lambda^2}{m_1^2}\, \bigg] \,\frac{\lambda^2}{m_1^2}\,
   \Psi(p,\frac{\lambda^2}{m_1^2})\,\nonumber\,\\
 & & \mbox{}
    + \frac{1 - {w^2}}{2\,{w^2}}\,
   \bigg[ 1 + \frac{3 - 5\,{w^2}}{4\,{w^2}}\,
      \frac{\lambda^2}{m_1^2} \,\bigg] \,
   \Phi(\frac{\lambda^2}{m_1^2})\,\nonumber\,\\
 & & \mbox{} +
   \frac{{{\left( 1 + w \right) }^2}\,
     \left( 1 - {w^2} \right) }{16\,{w^4}}\,
   \bigg[ \frac{3 - 5\,{w^2}}{{{\left( 1 + w \right) }^2}} +
     \frac{3\,{{\left( 1 - w \right) }^2}}{2\,w}\,\ln p \,
      \bigg] \,\frac{\lambda^4}{m_1^4}\,
   \ln \frac{\lambda^2}{m_1^2}\,\nonumber\,\\
 & & \mbox{} +
  \frac{1 - {w^2}}{2\,{w^2}}\,
   \bigg[ 1 + \frac{1 - {w^2}}{2\,w}\,\ln p \,\bigg] \,
   \frac{\lambda^2}{m_1^2}\,\ln \frac{\lambda^2}{m_1^2} +
  \bigg[\frac{3\,{{\left( 1 - {w^2} \right) }^3}}{64\,{w^5}}\,
   {{\ln p}^2}\,\bigg]\,\frac{\lambda^4}{m_1^4}\,\nonumber\,
   \\
 & & \mbox{} + \frac{1 - {w^2}}{4\,{w^2}}\,
   \bigg[ 1 + \frac{1 - {w^2}}{2\,w}\,\ln p\,
      \Big( 3 + \ln p \Big) \, \bigg] \,
   \frac{\lambda^2}{m_1^2} +
  \frac{1 - {w^2}}{4\,w}\,\ln p
   \label{F2oneloop}
\end{eqnarray}
where
\begin{eqnarray}
 \Phi(\xi) & = & \frac{1}{2}\sqrt{\xi^2-4\,\xi}
  \ln\left(\frac{\xi-\sqrt{\xi^2-4\,\xi}}
                {\xi+\sqrt{\xi^2-4\,\xi}}\right) \,,\label{Phi}\\[30mm]
\Psi(p,\xi) & = & \frac{1}{2}\ln^2\left(
  \frac{1}{2}\left[\xi-2+\sqrt{\xi^2-4\,\xi}\right]\right) \nonumber\\& &+
  \mbox{Li}_2\left(1+\frac{p}{2}
    \left[-2+\xi+\sqrt{\xi^2-4\,\xi}\right]\right) +
  \mbox{Li}_2\left(1+\frac{p}{2}
    \left[-2+\xi-\sqrt{\xi^2-4\,\xi}\right]\right) \label{Psi}\\& &+
  \left\{
    \begin{array}{ll}
     \displaystyle{
     -\frac{3}{2}\,\pi^2
     + 4\,\pi\arctan\left(\frac{\sqrt{4\,\xi-\xi^2}}{\xi}\right)
     + 2\,\pi\arctan\left(\frac{2\,p+\xi-2}
         {\sqrt{4\,\xi-\xi^2}}\right)\,,}
      & 0<\xi<4 \\
     \displaystyle{
     -\frac{1}{2}\,\pi^2 + i\,\pi\,\ln\Big((1-p)^2+p\,\xi\Big)\,,}
      & \xi>4
    \end{array} \right.\,.\nonumber
\end{eqnarray}
The QED normalization $\widehat{F}_1(0)=0$ is evidently adopted.
The evaluation of eq.~(\ref{mastervirt}) is tedious for arbitrary
$m_1$ and $m_2$
and will be presented in detail in~\cite{HKTfuture}.
In the limiting case $m_1\gg m_2$, however, the result is
drastically simplified:
\begin{eqnarray}
\mbox{Re}\,F_1 & = & f_1^{(2)}\,\ln^2\frac{{m_2^2}}{s} +
      f_1^{(1)}\,\ln\frac{{m_2^2}}{s} +
      f_1^{(0)}\,,\label{F1twoloop}  \\
\mbox{Re}\,F_2 & = & f_2^{(1)}\,\ln\frac{{m_2^2}}{s} +
      f_2^{(0)}   \label{F2twoloop} \,
\end{eqnarray}
where
\begin{eqnarray}
f_1^{(2)} & = &
\mbox{}\frac{1}{12}\,
  \bigg[\, \frac{1 + {w^2}}{2\,w}\,\ln p + 1 \,\bigg]\,,
  \label{fullvirtf12} \\
f_1^{(1)} & = &
\mbox{}\frac{1 + {w^2}}{6\,w}\,
   \bigg[ \mbox{Li}_2(1 - p) +
     \ln p\,\Big( \ln(1 + p) -
        \frac{1}{4}\,\ln p \Big)  - 3\,\zeta(2)
      \bigg] \,\nonumber\,\\
 & & \mbox{} +
  \frac{8 - 6\,w + 11\,{w^2}}{36\,w}\,\ln p +
  \frac{1}{3}\,\ln(1 + p) + \frac{11}{18}\,,
  \label{fullvirtf11} \\
f_1^{(0)} & = &
\mbox{}\frac{1 + {w^2}}{3\,w}\,
   \bigg[\, -\mbox{T}_3(1,0,w) +
     \mbox{T}_3(1,{1\over w},w) -
     \mbox{T}_3(1,w,{1\over w}) +
     2\,\ln w\,\mbox{T}_2(1,w) -
     \frac{\pi}{2}\,\mbox{T}^*_2(1,{1\over w}) -
     G\,\pi\,\nonumber\,\\
 & & \mbox{}\,
      \quad + \mbox{Li}_3({{1 - p}\over 2}) -
     \mbox{Li}_3(1 - p) -
     \mbox{Li}_3({{-1 + p}\over {2\,p}}) -
     \frac{1}{2}\,\mbox{Li}_3(p) +
     \frac{1}{2}\,\zeta(3)\,\nonumber\,
      \\
 & & \mbox{}\,\quad +
     \frac{1}{2}\,\Big( \mbox{Li}_2(p)\,
         \ln({{{w^4}\,
              \left( 1 - {w^2} \right) }\over 4}) -
        \mbox{Li}_2({p^2})\,\ln w \Big)  +
     \frac{1}{2}\,\ln 2\,\ln p\,
      \ln({{2\,p}\over
         {{{\left( 1 - p \right) }^2}}})\,
      \nonumber\,\\
 & & \mbox{}\,\quad +
     \frac{1}{2}\,\ln p\,\ln^2 w -
     \frac{1}{4}\,\ln^2 p\,\ln(1 - {p^2}) +
     \frac{5}{24}\,\ln^3 p +
     \Big( \frac{3}{2}\,\ln(4\,p) -
        2\,\ln(1 + p) \Big) \,\zeta(2) \,\bigg] \,
   \nonumber\,\\
 & & \mbox{} -
  \frac{8 + 11\,{w^2}}{18\,w}\,
   \bigg[\, \mbox{Li}_2(p) +
     \frac{1}{4}\,{{\ln p}^2} + \ln p\,\ln w +
     2\,\zeta(2) \,\bigg] \,\nonumber\,
   \\
 & & \mbox{} +
  \frac{1}{12}\,\ln^2({{1 - {w^2}}\over 4}) +
  \frac{131 - 132\,w + 134\,{w^2}}{216\,w}\,\ln p +
  \frac{11}{9}\,\ln(1 + p) +
  \frac{1}{3}\,\zeta(2) + \frac{67}{54}\,,
  \label{fullvirtf10} \\
f_2^{(1)} & = &
  -\,\frac{1 - {w^2}}{12\,w}\, \ln p \,,
  \label{fullvirtf21} \\
f_2^{(0)} & = &
\frac{1 - {w^2}}{6\,w}\,\bigg[\, -\mbox{Li}_2(1 - p) +
\frac{1}{4}\,\ln^2 p -
    \ln p \,\ln(1 + p) - \frac{25}{12}\,\ln p
    + 3\,\zeta(2) \,\bigg] \,
    \label{fullvirtf20}
\end{eqnarray}
and
\begin{eqnarray}
\mbox{T}_2(\eta,\xi) & \equiv &
  \int_0^1{\rm d}x\,\frac{\arctan(\xi\,x)}{x^2+\eta^2} \,,\nonumber\\
\mbox{T}_2^*(\eta,\xi) & \equiv &
  \int_0^1{\rm d}x\,\frac{\ln(x^2+\xi^2)}{x^2+\eta^2} \,,\nonumber\\
\mbox{T}_3(\eta,\xi,\chi) & \equiv &
  \int_0^1{\rm d}x\,\frac{\ln(x^2+\xi^2)\,\arctan(\chi\,x)}
       {x^2+\eta^2} \,,\label{Tdefs}\\
G & = & \int_0^1{\rm d}x\,\frac{\arctan(x)}{x}\,=\,
          0.915965594177219\ldots \quad\mbox{(Catalan's constant).}\nonumber
\end{eqnarray}
Again the functions $f_1^{(1,2)}$ and $f_2^{(1)}$ are closely
related to the very well-known logarithmically divergent and
constant pieces of the corresponding one loop
corrections for small $\lambda$~\cite{Schwinger}:
\begin{eqnarray}
\mbox{Re}\,\widehat{F}_1 & \stackrel{\lambda\to 0}{\longrightarrow} &
 \mbox{}
 6\,\bigg[\, f_1^{(2)} \, \bigg(
 \ln \frac{s}{\lambda^2} + \frac{5}{3} \, \bigg) - \frac{1}{2} \, f_1^{(1)}
 \,\bigg] \,,\nonumber\\
\mbox{Re}\,\widehat{F}_2 & \stackrel{\lambda\to 0}{\longrightarrow} &
 \mbox{}-3\,f_2^{(1)} \,.   \label{onetwovirt}
\end{eqnarray}
The similarity of these relations with eq.~(\ref{onetworeal}) is evident.
Interesting special cases are again the behaviour close to threshold
and for high energies. Let us discuss the former:
\begin{eqnarray}
\mbox{Re}\,F_1 & \stackrel{w\to 0}{\longrightarrow} &
\frac{1}{6\,w}\,\bigg[\, -3\,\ln \frac{m_2^2}{s} +
     6\,\ln w - 8 \,\bigg] \,\zeta(2) +
  \bigg[\, \frac{1}{2}\,\ln \frac{m_2^2}{s} + \ln 2 +
     \frac{1}{4} \,\bigg] \,\nonumber\,
   \\
 & & \mbox{} +
  \frac{1}{6}\,\bigg[\, -3\,\ln \frac{m_2^2}{s} +
     6\,\ln w - 11 \,\bigg] \,\zeta(2)\,w\,+{\cal O}(w^2)\,,
    \label{f1thresh} \\
\mbox{Re}\,F_2 & \stackrel{w\to 0}{\longrightarrow} &
\frac{1}{2\,w}\,\zeta(2) +
  \frac{1}{3}\,\bigg[\, \frac{1}{2}\,
      \ln \frac{m_2^2}{s} + \ln 2 + \frac{13}{12}
      \,\bigg]  - \frac{1}{2}\,\zeta(2)\,w\,+{\cal O}(w^2)\,.
  \label{f2thresh}
\end{eqnarray}
The Coulombic behaviour $\sim 1/w$ is evident from this result.
For $F_1$ the Coulomb singularity is
modified by the logarithmic factor $\ln(m_2^2/s)$ which is
responsible for the "running" of the coupling constant in the
${\cal O}(\alpha)$ result.
It is instructive to combine the vertex correction of ${\cal O}(\alpha)$
and ${\cal O}(\alpha^2)$ in the region close to threshold. As an
illustrative example we will examine the Dirac formfactor for this
case. The infrared divergent part of $\widehat{F}_1$ vanishes for
$w\rightarrow 0$ and therefore
\begin{eqnarray}
\lefteqn{
 \bigg(\frac{\alpha}{\pi}\bigg)\,\mbox{Re}\,\widehat{F}_1 +
 \bigg(\frac{\alpha}{\pi}\bigg)^2\,\mbox{Re}\,F_1\quad
 \stackrel{w\to 0}{\longrightarrow} }
    \nonumber\\
& &\mbox{}\,\frac{3}{2}\,\frac{\zeta(2)}{w}\,
  \bigg(\frac{\alpha}{\pi}\bigg)\,
   \bigg[ \, 1 + \bigg(\frac{\alpha}{\pi}\bigg)\frac{1}{3}\,
          \bigg(-\ln\frac{m_2^2}{s\,w^2} - \frac{8}{3}\,\bigg)\,\bigg]
  \nonumber\\
 & & \mbox{\hspace{4mm}}  + \frac{3}{2}\, \bigg(\frac{\alpha}{\pi}\bigg) \,
   \bigg[-1 + \bigg(\frac{\alpha}{\pi}\bigg) \frac{1}{3} \, \bigg(
         \ln\frac{4\,m_2^2}{s} + \frac{1}{2} \, \bigg) \, \bigg] \, +
   {\cal O}(w)\,.
   \label{oneplustwoF1}
\end{eqnarray}
The fine structure constant $\alpha$, defined at vanishing
momentum transfer, is related to the $\overline{\mbox{MS}}$ coupling
constant at subtraction point $\mu^2$ by
\begin{equation}
\alpha = \alpha_{\overline{\mbox{\tiny MS}}}(\mu^2)\bigg(
   1+\frac{\alpha_{\overline{\mbox{\tiny MS}}}(\mu^2)}{\pi}\,\frac{1}{3}\,
   \ln\frac{m_2^2}{\mu^2}\,\bigg) +
   {\cal O}(\alpha^3_{\overline{\mbox{\tiny MS}}})\,.
   \label{alphaphystorun}
\end{equation}
At this point it becomes obvious that the natural scale for
$\alpha_{\overline{\tiny\mbox{MS}}}$ in the threshold region is given
by the nonrelativistic momentum $\mu^2 = s\,w^2$ as far as the
$1/w$~terms are concerned. [This holds true as long as
$w \, {\scriptscriptstyle\stackrel{\scriptscriptstyle>}{\sim}} \, \alpha$.
Below this value the approximations used in this work are no longer
applicable.] For the correction resulting from transverse photon
exchange, which are not enhanced by $1/w$, the scale $\mu^2 = s/4$ is
appropriate.
This suggests the following form of the Dirac formfactor in the
threshold region
\begin{eqnarray}
\lefteqn{
 \bigg(\frac{\alpha}{\pi}\bigg)\,\mbox{Re}\,\widehat{F}_1 +
 \bigg(\frac{\alpha}{\pi}\bigg)^2\,\mbox{Re}\,F_1\quad
 \stackrel{w\to 0}{\longrightarrow} } \nonumber\\
 & & \mbox{}
  \frac{3}{2}\,\frac{\zeta(2)}{w}\,
  \bigg(\frac{\alpha_{\overline{\tiny\mbox{MS}}}(s\,w^2)}{\pi}\bigg)
 \bigg[ \, 1 - \bigg(\frac{\alpha}{\pi}\bigg) \frac{8}{9} \, \bigg]
 \nonumber \\
 & & \hskip 3mm \mbox{}\,
  - \frac{3}{2}\,
 \bigg(\frac{\alpha_{\overline{\tiny\mbox{MS}}}(s/4)}{\pi}\bigg)\,
 \bigg[ \, 1 - \bigg(\frac{\alpha}{\pi}\bigg)\frac{1}{6}\,\bigg] \, + \,
 {\cal O}(w)
\end{eqnarray}
where the scale in the ${\cal O}(\alpha^2)$ term is not yet determined.
It is clear that the virtual corrections will dominate the rate
\pagebreak[3]
close to the threshold.
\par
For high energies, on the other hand, one finds
\begin{eqnarray}
\mbox{Re}\,F_1 & \stackrel{\frac{m_1^2}{s}\to 0}{\longrightarrow} &
\frac{1}{12}\,\ln^2 \frac{m_2^2}{s}\,
   \Big( \ln x + 1 \Big)  +
  \frac{1}{12}\,\ln \frac{m_2^2}{s}\,
   \bigg( -\ln^2 x + \frac{13}{3}\,\ln x -
     8\,\zeta(2) + \frac{22}{3} \bigg) \,
   \nonumber\,\\
 & & \mbox{} +
  \frac{1}{36}\,\ln^3 x - \frac{13}{72}\,\ln^2 x +
  \bigg( \frac{1}{3}\,\zeta(2) +
     \frac{133}{216} \bigg) \,\ln x -
  \frac{1}{3}\,\zeta(3) - \frac{16}{9}\,\zeta(2) +
  \frac{67}{54}\,\nonumber\,\\
 & & \mbox{} +
  x\,\bigg[\, \frac{1}{6}\,\ln^2 \frac{m_2^2}{s} +
     \frac{1}{6}\,\ln \frac{m_2^2}{s}\,
      \left( \ln x + \frac{13}{3} \right) \,
      \nonumber\,\\
 & & \mbox{}\,
      \quad\quad\, - \frac{1}{12}\,\ln^2 x +
     \frac{49}{36}\,\ln x - \frac{4}{3}\,\zeta(2) +
     \frac{115}{108} \,\bigg] \,
    +{\cal O}(x^2)\,,
    \label{F1high}\\
\mbox{Re}\,F_2 & \stackrel{\frac{m_1^2}{s}\to 0}{\longrightarrow} &
x\,\bigg[\, -\, \frac{1}{3}\,\ln \frac{m_2^2}{s}\,
        \ln x   + \frac{1}{6}\,\ln^2 x -
     \frac{25}{18}\,\ln x + \frac{4}{3}\,\zeta(2)
      \,\bigg] \,
    +{\cal O}(x^2)\,.
    \label{F2high}
\end{eqnarray}
For the special case $m_1=m_2=m$ the integrals~(\ref{mastervirt}) lead to
particularly simple results
\begin{eqnarray}
\mbox{Re}\,F_1 & = &
\frac{1}{12\,{w^2}}\,
  \bigg[\, \frac{\left( -1 + 2\,{w^2} \right) \,
       \left( 9 - 6\,{w^2} + 5\,{w^4} \right) }{
      24\,{w^3}}\,\ln^3 p +
    \frac{31 - 23\,{w^2} + 30\,{w^4}}{12\,{w^2}}\,
     \ln^2 p \,\nonumber\,\\
 & & \mbox{}\,
     \quad + \bigg( \frac{
        267 - 238\,{w^2} + 236\,{w^4}}{18\,w} +
       \frac{\left( 1 - 2\,{w^2} \right) \,
          \left( 9 - 6\,{w^2} + 5\,{w^4} \right) }{
         2\,{w^3}}\,\zeta(2) \bigg) \,\ln p\,
     \nonumber\,\\
 & & \mbox{}\,\quad +
    \frac{9 - 21\,{w^2} - 10\,{w^4}}{{w^2}}\,
     \zeta(2) + \frac{147 + 236\,{w^2}}{9} \,\bigg]\,,
    \label{fullF1equal}\\
\mbox{Re}\,F_2 & = &
    \label{fullF2equal}
\frac{1 - {w^2}}{4\,{w^2}}\,
  \bigg[\, \frac{{{\left( 1 - {w^2} \right) }^2}}{
      8\,{w^3}}\,\ln^3 p -
    \frac{11 - 9\,{w^2}}{12\,{w^2}}\,\ln^2 p\,
     \nonumber\,\\
 & & \mbox{}\,
     \quad - \Big( \frac{93 - 68\,{w^2}}{18\,w} +
       \frac{3\,{{\left( 1 - {w^2} \right) }^2}}{
         2\,{w^3}}\,\zeta(2) \Big) \,\ln p -
    \frac{17}{3} + \frac{-3 + 5\,{w^2}}{{w^2}}\,
     \zeta(2) \,\bigg] \,
\end{eqnarray}
with the high energy expansion
\begin{eqnarray}
\mbox{Re}\,F_1 & \stackrel{\frac{m^2}{s}\to 0}{\longrightarrow} &
\frac{1}{36}\,\ln^3 x + \frac{19}{72}\,\ln^2 x +
  \frac{1}{3}\,\Big( \frac{265}{72} - \zeta(2) \Big) \,\ln x -
  \frac{11}{6}\,\zeta(2) + \frac{383}{108}\,\nonumber\\
 & & \mbox{} +
  x\,\bigg[ \frac{5}{4}\,\ln^2 x + \frac{49}{12}\,\ln x -
     \frac{5}{3}\,\zeta(2) + \frac{853}{108} \bigg] \,
    +{\cal O}(x^2)\,,
    \label{F1equalhigh} \\
\mbox{Re}\,F_2 & \stackrel{\frac{m^2}{s}\to 0}{\longrightarrow} &
x\,\bigg[\, - \frac{1}{6}\,\ln^2 x   - \frac{25}{18}\,\ln x +
     2\,\zeta(2) - \frac{17}{3} \,\bigg] \,
     +{\cal O}(x^2)\,.
     \,\label{F2equalhigh}
\end{eqnarray}
The logarithmically enhanced and the constant parts are
in agreement with~\cite{BURGERS}.
\par
Finally, the ${\cal O}(\alpha^2)$ contribution of the virtual corrections
to the rate for the case $m_1\gg m_2$ is given by
\begin{eqnarray}
\label{deltavirt}
\delta R_V & = &
 \left(\frac{\alpha}{\pi}\right)^2\,\varrho^V \,,\nonumber\\
\varrho^V & = &
   w\,(3-w^2)\,\Big(\mbox{Re}\,F_1+\mbox{Re}\,F_2\Big)
   +w^3\,\mbox{Re}\,F_2 \,.
\end{eqnarray}
\par
\vspace{1cm}
\noindent
{\em 4.~The total rate}\\
Combining real and virtual radiation one thus arrives at
\begin{eqnarray}
\label{realplusvirt}
\varrho^R + \varrho^V & = &
   -\,\frac{1}{3}\,W\,\ln\frac{m_2^2}{s} + f_{R}^{(0)} +
   w\,(3-w^2)\,\left(f_1^{(0)}+f_2^{(0)}\right)+
   w^3\,f_2^{(0)} \,.
\end{eqnarray}
The quadratic logarithm in $m_2^2/s$ from the real and the virtual
radiation cancel. A linear logarithm, however, remains. Its origin can be
easily understood  through the running of the coupling constant
$\alpha$. The prefactor $W$ is identical to the correction function of
${\cal O}(\alpha)$ derived by Schwinger~\cite{Schwinger}.
Therefore the expression for the rate for the inclusive
$f_1 \bar f_1$ final state including ${\cal O}(\alpha)$ photonic
corrections plus photonic ${\cal O}(\alpha^2)$ corrections due
to one light fermion with mass $m_2$ reads
\begin{eqnarray}
R & = & \frac{1}{2}\,w\,(3-w^2) +
  \left(\frac{\alpha}{\pi}\right)\,W \nonumber\\ & &+
 \left(\frac{\alpha}{\pi}\right)^2\,\bigg[
   -\frac{1}{3}\,W\,\ln\frac{m_2^2}{s} + f_{R}^{(0)} +
   w\,(3-w^2)\,\left(f_1^{(0)}+f_2^{(0)}\right)+
   w^3\,f_2^{(0)} \,\bigg] \nonumber
\end{eqnarray}
where
\begin{eqnarray}
W & = & - 3 \bigg[ f_R^{(1)} + w(3-w^2) \bigg(
 f_1^{(1)} + f_2^{(1)}\bigg) + w^3\,f_2^{(1)} \bigg] = \nonumber\\
 & = & \mbox{}
\frac{\left( 3 - {w^2} \right) \,\left( 1 + {w^2} \right) }{2
    }\,\bigg[\, 2\,\mbox{Li}_2(p) + \mbox{Li}_2({p^2}) +
     \ln p\,\Big( 2\,\ln(1 - p) + \ln(1 + p) \Big)
      \,\bigg] \,\nonumber\,\\
 & & \mbox{} -
  w\,( 3 - {w^2} ) \,
   \Big( 2\,\ln(1 - p) + \ln(1 + p) \Big)  +
  \frac{\left( -1 + w \right) \,
     \left( 33 - 39\,w - 17\,{w^2} + 7\,{w^3} \right) }{16}\,
   \ln p\,\nonumber\,\\
 & & \mbox{} +
  \frac{3\,w\,\left( 5 - 3\,{w^2} \right) }{8}\,.
  \label{deltaoneloop}
\end{eqnarray}
[Note that the massive quark is not
accounted for, consistent with the fact that virtual heavy fermion loops are
not considered in eq.~(\ref{fullreal}). Adding virtual corrections
eq.~(\ref{fullF1equal},~\ref{fullF2equal}) and the
real radiation, e.g.~based on a numerical evaluation of eq.~(\ref{masterreal})
one would thus include ``double bubble'' diagrams with two massive
fermions. This will be done in~\cite{HKTfuture}.]
\par
Relating again the fine structure constant $\alpha$ to the
$\overline{\mbox{MS}}$ coupling $\alpha_{\overline{\mbox{\tiny MS}}}$
at the scale $\mu^2$, the mass singularities
disappear and one finds
\begin{eqnarray}
R & = & \frac{1}{2}\,w\,(3-w^2) +
  \bigg(\frac{\alpha_{\overline{\tiny\mbox{MS}}}(\mu^2)}{\pi}\bigg)\,W
  \nonumber\\ & &+\,
 \bigg(\frac{\alpha_{\overline{\mbox{\tiny MS}}}(\mu^2)}{\pi}\bigg)^2
 \bigg[\, -\frac{1}{3}\,W\,\ln\frac{\mu^2}{s}  +
   f_{R}^{(0)} +
   w\,(3-w^2)\,\left(f_1^{(0)}+f_2^{(0)}\right)+
   w^3\,f_2^{(0)} \,\bigg] \,.
   \label{totalrate}
\end{eqnarray}
The behaviour close to threshold for the choice $\mu^2=s$ is
easily read off from eq.~(\ref{totalrate}):
\begin{eqnarray}
R & \stackrel{w\to 0}{\longrightarrow} &
\frac{3}{2}\,w\,\nonumber\,\\
 & & \mbox{} +
  \bigg(\frac{\alpha_{\overline{\tiny\mbox{MS}}}(s)}{\pi}\bigg)\,
   \bigg[\, \frac{9}{2}\,\zeta(2) - 6\,w \,\bigg] \,\nonumber\,\\
 & & \mbox{} +
  {{\bigg(\frac{\alpha_{\overline{\tiny\mbox{MS}}}(s)}{\pi}\bigg)}^2}\,
   \bigg[\, \bigg( 3\,\ln w - \frac{5}{2} \bigg) \,\zeta(2) +
     \bigg( 4\,\ln 2 + \frac{11}{6} \bigg) \,w \,\bigg]\,
     +{\cal O}(w^2)\,.
     \label{Rthresh}
\end{eqnarray}
The discussion following eq.~(\ref{oneplustwoF1}) applies equally well to this
formula, since real radiation vanishes close to threshold.
\par
\begin{figure}[htb]
\begin{center}
\leavevmode
\epsfxsize=5cm
\epsffile[220 420 420 550]{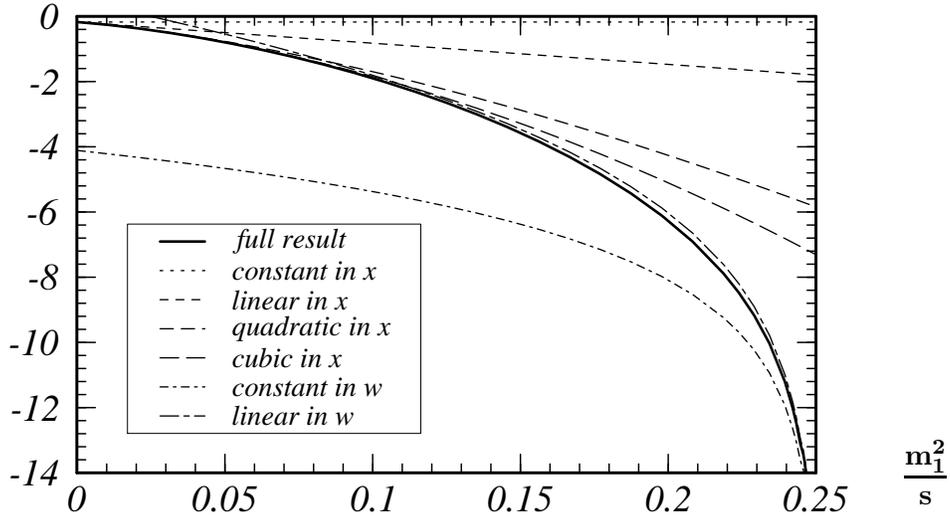}
\vskip 30mm  $\displaystyle{\mbox{\hspace{12.5cm}}\bf\frac{m_1^2}{s}}$
\vskip 5mm
\caption{\label{realplusvirtfig} {\em
${\cal O}(\alpha^2_s(s))$ correction to the inclusive
production rate $R_{f_1 \bar f_1}$ based on the exact result
(solid line) and approximations described in the text. }}
\end{center}
\end{figure}

In the high energy region one finds
\begin{eqnarray}
R & \stackrel{\frac{m_1^2}{s}\to 0}{\longrightarrow} &
1 - 6\,{x^2} - 8\,{x^3}\,\nonumber\,\\
 & & \mbox{} +
  \bigg(\frac{\alpha_{\overline{\tiny\mbox{MS}}}(s)}{\pi}\bigg)\,
   \bigg\{ \frac{3}{4} + 9\,x + {x^2}\,
      \bigg[\, -18\,\ln x + \frac{15}{2} \,\bigg]  -
     {x^3}\,\bigg[\, \frac{116}{3}\,\ln x + \frac{188}{9} \,\bigg]  \bigg\} \,
   \nonumber\,\\
 & & \mbox{} +
  {{\bigg(\frac{\alpha_{\overline{\tiny\mbox{MS}}}(s)}{\pi}\bigg)}^2}\,
   \bigg\{ \zeta(3) - \frac{11}{8} -
     \frac{13}{2}\,x\,\nonumber\,\\
 & & \mbox{}\,\qquad +
     {x^2}\,\bigg[\, -3\,\ln^2 x + \frac{27}{2}\,\ln x - 4\,\zeta(3) -
        18\,\zeta(2) - \frac{35}{6}\, \bigg] \,\nonumber\,\\
 & & \mbox{}\,
      \qquad + {x^3}\,\bigg[\, -8\,\ln^2 x + \frac{752}{27}\,\ln x -
        \frac{304}{9}\,\zeta(2) + \frac{1282}{81} \,\bigg]  \bigg\} \,
    +{\cal O}(x^4)\,.
    \label{Rhigh}
\end{eqnarray}
In Fig.~\ref{realplusvirtfig} the comparison between the
exact ${\cal O}(\alpha^2)$ correction for $\mu^2=s$ (solid line),
threshold approximations (dashed dotted lines)
and high energy expansions (dashed lines) is performed.
Eq.~(\ref{Rhigh}) provides an important consistency check
on our result. It is straightforward to relate pole and
$\overline{\mbox{MS}}$ definition for the remaining fermion mass $m_1$
taking again into account in ${\cal O}(\alpha^2)$
only the contribution from one light virtual fermion:
\begin{eqnarray}
  m_1^2 & = &
\overline{m}_1^2(\mu^2)\,\bigg\{ 1 +
    \bigg(\frac{\alpha_{\overline{\tiny\mbox{MS}}}(\mu^2)}{
\pi}\bigg)\,
     \bigg[ -\frac{3}{2}\,\ln \frac{\overline{m}_1^2(\mu^2)}{\mu^2} +
        2 \,\bigg] \,\nonumber\,\\
 & & \mbox{}\qquad\qquad\,\,\, +
    \bigg(\frac{\alpha_{\overline{\tiny\mbox{MS}}}(\mu^2)}{
\pi}\bigg)^2\,
     \bigg[ -\frac{1}{4}\,\ln^2 \frac{\overline{m}_1^2(\mu^2)}{\mu^2}
         + \frac{13}{12}\,\ln \frac{\overline{m}_1^2(\mu^2)}{\mu^2} -
         \zeta(2) -
       \frac{71}{48} \,\bigg]  \bigg\}\,.
   \label{mpoltorun}
\end{eqnarray}
Replacing the pole mass by the running mass at the scale
$\mu^2=s$ the logarithmic factor of the $\overline{m}^2$ term disappears
as expected from general considerations. The structure of the
logarithms of the $\overline{m}^2$ and $\overline{m}^4$
terms coincides with
the expectations from~\cite{CKfirst, CKlast}. In fact, after replacing
the abelian factors by the proper SU(3)-coefficients one obtains
\begin{eqnarray}
R & = &
1 + \bigg(\frac{\alpha_s(s)}{\pi}\bigg) +
  n_f\,\bigg(\frac{\alpha_s(s)}{\pi}\bigg)^2\,
   \bigg[\, \frac{2}{3}\,\zeta(3) - \frac{11}{12} \,\bigg] \,\nonumber\,
   \\
 & & \mbox{} + \frac{\overline{m}_1^2(s)}{s}\,
   \bigg\{ 12\,\bigg(\frac{\alpha_s(s)}{\pi}\bigg) -
     \frac{13}{3}\,n_f\,\bigg(\frac{\alpha_s(s)}{\pi}\bigg)^2 \,\bigg\} \,
   \nonumber\,\\
 & & \mbox{} +
  \frac{\overline{m}_1^4(s)}{s^2}\,
   \bigg\{ -6 - 22\,\bigg(\frac{\alpha_s(s)}{\pi}\bigg) +
     n_f\,\bigg(\frac{\alpha_s(s)}{\pi}\bigg)^2\,
      \bigg[\, \frac{1}{3}\,\ln \frac{\overline{m}_1^2(s)}{s} -
        \frac{8}{3}\,\zeta(3) - 4\,\zeta(2) + \frac{143}{18} \,\bigg]
        \bigg\} \,
   \nonumber\,\\
 & & \mbox{} +
  \frac{\overline{m}_1^6(s)}{s^3}\,
   \bigg\{ -8 + \bigg(\frac{\alpha_s(s)}{\pi}\bigg)\,
      \bigg[\, - \frac{32}{9}\,\ln \frac{\overline{m}_1^2(s)}{s}
      - \frac{2480}{27} \,\bigg] \,\nonumber\,\\
 & & \mbox{}\,
      \qquad\qquad + n_f\,\bigg(\frac{\alpha_s(s)}{\pi}\bigg)^2\,
      \bigg[\, - \frac{4}{3}\,\ln^2 \frac{\overline{m}_1^2(s)}{s} +
        \frac{100}{81}\,\ln \frac{\overline{m}_1^2(s)}{s} -
        \frac{176}{27}\,\zeta(2) + \frac{8315}{243} \,\bigg]  \bigg\}
\label{rqcd}
\end{eqnarray}
where now the number of light fermions, $n_f$, is displayed explicitly.
The relation to the $n_f$ dependent terms of eq.~(27) in~\cite{CKlast} is
evident.
\par
\vspace{1cm}\noindent
{\em 5.~Summary}\\
The rate for the production of a pair of massive fermions in $e^+ e^-$
annihilation plus real and virtual radiation of a pair of light fermions
has been calculated analytically. This result, together with~\cite{HJKT2}
can be considered as a first step towards the evaluation of the production
cross section for heavy fermions in ${\cal O}(\alpha^2)$. The expansion
of the result for energies close to threshold and for high energies
and subsequent comparisons with earlier asymptotic formulas
provide important cross checks. The transition to the $\overline{\mbox{MS}}$
scheme leads to interesting
insights into the proper scale of the coupling constant.
\par
\vspace{1cm}\noindent
{\em Acknowledgement:}
We would like to thank K.~Chetyrkin and M.~Je\.zabek for helpful
discussions.
\vskip 1.5cm
\sloppy
\raggedright
\def\app#1#2#3{{\it Act. Phys. Pol. }{\bf B #1} (#2) #3}
\def\apa#1#2#3{{\it Act. Phys. Austr.}{\bf #1} (#2) #3}
\def\lhc{Proc. LHC Workshop, CERN 90-10}
\def\npb#1#2#3{{\it Nucl. Phys. }{\bf B #1} (#2) #3}
\def\plb#1#2#3{{\it Phys. Lett. }{\bf B #1} (#2) #3}
\def\prd#1#2#3{{\it Phys. Rev. }{\bf D #1} (#2) #3}
\def\pR#1#2#3{{\it Phys. Rev. }{\bf #1} (#2) #3}
\def\prl#1#2#3{{\it Phys. Rev. Lett. }{\bf #1} (#2) #3}
\def\prc#1#2#3{{\it Phys. Reports }{\bf #1} (#2) #3}
\def\cpc#1#2#3{{\it Comp. Phys. Commun. }{\bf #1} (#2) #3}
\def\nim#1#2#3{{\it Nucl. Inst. Meth. }{\bf #1} (#2) #3}
\def\pr#1#2#3{{\it Phys. Reports }{\bf #1} (#2) #3}
\def\sovnp#1#2#3{{\it Sov. J. Nucl. Phys. }{\bf #1} (#2) #3}
\def\jl#1#2#3{{\it JETP Lett. }{\bf #1} (#2) #3}
\def\jet#1#2#3{{\it JETP Lett. }{\bf #1} (#2) #3}
\def\zpc#1#2#3{{\it Z. Phys. }{\bf C #1} (#2) #3}
\def\ptp#1#2#3{{\it Prog.~Theor.~Phys.~}{\bf #1} (#2) #3}
\def\nca#1#2#3{{\it Nouvo~Cim.~}{\bf #1A} (#2) #3}
{\elevenbf\noindent  References \hfil}
\vglue 0.4cm


\begin{thebibliography}{9}
\bibitem{BFK}V.N. Ba\u{\i}er, V.S. Fadin and V.A. Khoze,
  {\it Soviet Physics} \jet{23}{1966}{104}.
\bibitem{Schwinger}J.S. Schwinger, {\it Particles, sources and fields},
  Addison-Wesley Publishing Company, Inc.(1970/73) and refs.~therein.
\bibitem{as2}K.G. Chetyrkin, A.L. Kataev and F.V. Tkachov,
  \plb{85}{1979}{277}; \\
  M. Dine and J. Sapirstein, \prl{43}{1979}{668};\\
  W. Celmaster and R.J. Gonsalves, \prl{44}{1980}{560}.
\bibitem{as3}S.G. Gorishny, A.L. Kataev and S.A. Larin,
  \plb{259}{1991}{144}; \\
  L.R. Surguladze and M.A. Samuel, \prl{66}{1991}{560}
  and 2416 (Erratum).
\bibitem{CKfirst}K.G. Chetyrkin and J.H. K\"uhn,
  \plb{248}{1990}{359}.
\bibitem{CKlast}K.G. Chetyrkin and J.H. K\"uhn, \npb{432}{1994}{337}.
\bibitem{Lewin}L.~Lewin, {\it Polylogarithms and associated functions},
  Elsevier North Holland, Inc., 1981.
\bibitem{HKTfuture}A.H. Hoang, J.H. K\"uhn and T. Teubner, in
  preparation.
\bibitem{HJKT2}A.H.~Hoang, M.~Je\.zabek, J.H.~K\"uhn and T.~Teubner,
  \plb{338}{1994}{330}.
\bibitem{KKKS} B.A. Kniehl, M. Krawczyk, J.H. K\"uhn and
R.G. Stuart, \plb{209}{1988}{337}.
\bibitem{HJKT1} A.H.~Hoang, M.~Je\.zabek, J.H.~K\"uhn and T.~Teubner,
  \plb{325}{1994}{495} and \plb{327}{1994}{439} (Erratum).
\bibitem{BURGERS}G.J.H. Burgers, \plb{164}{1985}{167}.
\end{thebibliography}
\end{document}